\documentclass[aps,pre,twocolumn,showpacs,preprintnumbers,a4paper]{revtex4} 

\usepackage{graphics}
\usepackage{epsf}
\usepackage{epsfig}

\newcommand{\tindex}[1]{{\scriptstyle \text{#1}}}

\newcommand{\HC}{H_\tindex{c}}

\begin{document} 

\flushbottom

\title{The depinning transition of a driven interface in 
       the random-field Ising model
       around the upper critical dimension}

\author{L. Roters}
\email{lars@thp.uni.duisburg.de}
\affiliation{
Theoretische Tieftemperaturphysik,
Gerhard-Mercator-Universit\"at, 
Lotharstr.~1, 
47048 Duisburg, Germany}

\author{S. L\"ubeck}
\email{sven@thp.uni.duisburg.de}
\affiliation{
Weizmann Institute of Science
Department of Complex Physics,
76100 Rehovot, Israel
}
\affiliation{
Theoretische Tieftemperaturphysik,
Gerhard-Mercator-Universit\"at, 
Lotharstr.~1, 
47048 Duisburg, Germany}

\author{K.\,D. Usadel}
\email{usadel@thp.uni.duisburg.de}
\affiliation{
Theoretische Tieftemperaturphysik,
Gerhard-Mercator-Universit\"at, 
Lotharstr.~1, 
47048 Duisburg, Germany}

\date\today

\begin{abstract}
  We investigate the depinning transition for
  driven interfaces in the random-field Ising model 
  for various dimensions.
  We consider the order parameter as a function of the 
  control parameter (driving field) and examine the effect
  of thermal fluctuations.
  Although thermal fluctuations drive the system away from
  criticality the order parameter obeys a certain scaling law
  for sufficiently low temperatures and the corresponding
  exponents are determined.
  Our results suggest that the so-called upper critical dimension
  of the depinning transition is five and that the systems 
  belongs to the universality class of the quenched Edward-Wilkinson
  equation.
\end{abstract}


\preprint{\textit{accepted for publication in Phys.~Rev.~E}}

\maketitle
\section{Introduction}

Driven Interfaces in quenched disordered systems
display with increasing driving force a transition
from a pinned interface to a moving interface~(see
e.g.~\cite{KARDAR_2} and references therein).  
This so-called depinning transition is caused
by a competition of the driving force and the 
quenched disorder.
The first one tends to move the interface
whereas the latter one hinders the movement.
Depinning transitions are observed in a large variety
of physical problems, such as fluid invasion in porous
materials (see, for instance, Sec.~6.2 in \cite{HINRICHSEN_1} and
references therein), depinning of 
charge density waves~\cite{FISHER_1,FISHER_2}, 
impurity pinning of flux-line in type-II
superconductors~\cite{BLATTER_1},  
contact lines~\cite{DEGENNES_1} as well as in 
field driven ferromagnets,
where the interface separates regions of opposite
magnetizations~\cite{BRUINSMA}.

A well established model to investigate the depinning 
transition in disordered ferromagnets is the
driven random-field Ising model (RFIM)~(see for
instance~\cite{JI_2,DROSSEL_1,BRUINSMA,AMARAL_1,AMARAL_2}).
Here, the disorder induces some effective energy barriers
which suppress the interface motion.
A magnetic driving field~$H$ reduces these energy barriers 
but they vanish only if the driving field
exceeds the critical value $H_{\text c}$.
The transition from the pinned to the moving
interface can be described as a continuous
phase transition and its velocity~$v$  
is interpreted as the order parameter.
Without thermal fluctuations ($T=0$) the field dependence of the
velocity obeys the power-law behavior
\begin{equation}
v(h,T=0) \; \sim \;  h^\beta
\label{eq:v_H_01}
\end{equation}
for $h>0$, where $h$ denotes the reduced driving field
$h=H/H_{\text c}-1$.

The depinning transition is destroyed in the 
presence of thermal fluctuations ($T>0$)
which may provide the energy needed to overcome 
local energy barriers.
Although thermal fluctuations drive the system away 
from criticality the order parameter obeys certain
scaling laws and for sufficiently low temperatures
the order parameter can be described 
as a generalized homogenous function~\cite{ROTERS_1}
\begin{equation}
v(h,T) \; = \; \lambda \,\, {\tilde v} 
(\lambda^{-1/\beta} h, \lambda^{-\psi} T),
\label{eq:v_scal_01}
\end{equation}
similar to usual equilibrium second order phase transitions.
Setting $\lambda^{-1/\beta}=1$ one recovers
Eq.\,(\ref{eq:v_H_01}) for zero temperature.
Choosing $\lambda^{-\psi}\,T=1$ one gets for
the interface velocity at the critical field~$H_{\text c}$
\begin{equation}
  v( h = 0, T )  \sim  T^{1/\psi}.
  \label{eq:v_T_01}
\end{equation}
This power-law behavior was observed in two- and 
three-dimensional simulations of the driven 
RFIM~\cite{NOWAK_1,ROTERS_1}
as well as in charge density waves in computer simulations and
mean-field calculations~\cite{FISHER_2,MIDDLETON_1}. 

Furthermore, thermal fluctuations cause a creep motion of the 
interface for small driving fields ($H\ll H_{\text c}$)
characterized by an Arrhenius like behavior of the velocity
\cite{IOFFE_1,NATTERMANN_4}.
Recently, this creep motion was observed in experiments
considering magnetic domain wall motion in thin films composed of Co
and Pt layers~\cite{LEMERLE_1}, in renormalization group
calculations~\cite{CHAUVE_1,CHAUVE_2} regarding the Edwards-Wilkinson
equation with quenched disorder (QEW), as well as in numerical
simulations of the RFIM~\cite{ROTERS_4}.  

In equilibrium physics a scaling ansatz according
to~Eq.\,(\ref{eq:v_scal_01}) usually describes 
the order parameter as a function of its control parameter and of its
conjugated field.
Although Eq.\,(\ref{eq:v_scal_01}) can be applied to the depinning
transition, i.e. the temperature is a
relevant scaling field,~$T$ is not conjugated to the order parameter.
The conjugated field would support the interface motion independent of
its strength.
But strong thermal fluctuations destroy the interface instead
to support the interface motion.
Therefore, one has to interpret the value of the thermal
exponent~$\psi$ carefully.
For instance, it is not clear wether the obtained values of~$\psi$ are
a characteristic feature of the whole universality class of the
depinning transition or just a characteristic feature of the
particular considered RFIM. 
This point could be important for the interpretation of experiments
which naturally take place at finite temperatures.

In this paper we reinvestigate the interface dynamics of
the driven RFIM and focus our attention to higher
dimensions $d\ge 3$.
In particular we consider the scaling behavior at
the critical point and determine the exponents
$\beta$ and $\psi$.
Our results suggest that the so-called upper 
critical dimension of the depinning transition of
the driven RFIM is $d_{\text c}=5$.
Above this dimension the scaling behavior is characterized
by the mean-field exponents.
We compare our results with those obtained from a
renormalization group approach of the 
quenched Edward-Wilkinson equation
which is expected to be in the same universality
class as the driven RFIM.
A summary is given at the end.
\section{Model and simulations}

We consider the depinning transition
RFIM on cubic lattices of linear size $L$ in higher dimensions ($d\ge 3$).
The Hamiltonian of the RFIM is given by
\begin{equation}
  {\cal H} =
  -\frac{J}{2}\, \sum_{\langle i,j \rangle} S_i \, S_j
  -H \, \sum_{i} S_i
  - \sum_{i} h_i\,S_i
  \mbox{ ,}
\end{equation}
where the first term characterizes the exchange interaction 
of neighboring spins ($S_i=\pm 1$).
The sum is taken over all pairs of neighbored spins.
The spins are coupled to a homogenous driving field $H$ as well
as to a quenched random-field $h_i$ with $\langle h_i \rangle = 0$
and $\langle h_i h_j \rangle \propto \delta_{ij}$.
The random field is assumed to be uniformly distributed, 
i.e., the probability $p$ that the random field at site~$i$
takes some value $h_i$ is given by
\begin{equation}
  p(h_i) = 
  \left\{
    \begin{array}{ccl}
      (2\Delta)^{-1} & \;{\rm for}\; &|h_i| < \Delta\\
      0 & & \mbox{otherwise.}
    \end{array}
  \right.
\end{equation}
Using antiperiodic boundary conditions an
interface is induced into the system 
which can be driven by the field $H$
(see~\cite{ROTERS_1} for details).
A Glauber dynamics with random sequential 
update and heat-bath transition probabilities
is applied to simulate the interface motion
(see for instance~\cite{BINDER_1}). 

Due to this algorithm not only spins
adjacent to the interface but throughout the whole system can
flip with a finite probability at temperatures $T>0$.
In general, this may cause nucleation which always starts with an
isolated spin-flip.
The minimum energy required for an isolated spin-flip is
$\Delta E=2(z\,J-H-\Delta)$
with $z=2^d$ nearest neighbors on a bcc lattice.
Although the corresponding spin-flip probability
$1/[1+\exp(\Delta E/T)]$ is small for the considered temperatures
($T\le 0.2$), isolated spin-flips are possible and occur. But we
observed that these spin-flips  are unstable in our simulations,
i.e., an isolated spin will flip back in the next update step.
Thus the originally induced interfaces is stable during the
simulations.

The moving interface corresponds to a magnetization~$M$ which increases
in time~$t$ (given in Monte Carlo step per spin).
The interface velocity, which is the basic quantity in our
investigations, is obtained from the time dependence of the
magnetization 
$v=\langle {\text d}M/{\text d}t \rangle$ where $\langle\ldots \rangle$
denotes an appropriate disorder average.
Starting with a flat interface we performed a sufficiently
number of updates until the system reaches after a transient
regime the steady state which is characterized by a constant
average interface velocity.

\begin{figure}[b]
  \includegraphics[width=8.6cm,angle=0]{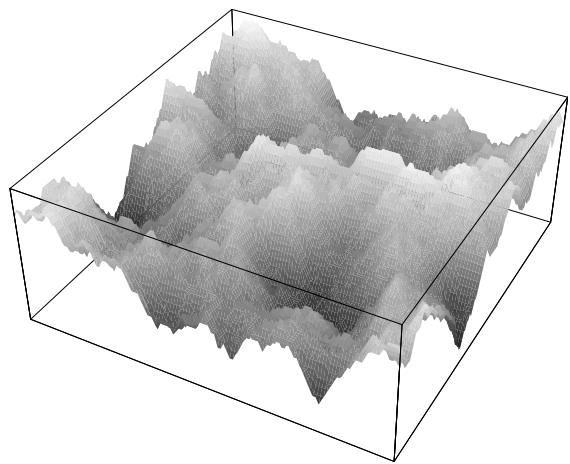}
  \caption{
    Snapshot of a moving interface
    at $T=0$ for $L=128$, $\Delta=1.7$, and $H=1.37$ in $d=3$.
    In order to show the details we stretched 
    the interface in the vertical direction by a factor 15. 
    }
 \label{fig:snapshot}
\end{figure}
As pointed out in previous works~\cite{NOWAK_1,ROTERS_1} 
an appropriate choice of the interface orientation is needed in
order to recover that the interface moves for arbitrarily small 
driving field in the absence of disorder.
An appropriate choice is to consider the interface 
motion along the diagonal direction of a simple cubic 
lattice.
For $d\ge 3$ an alternative is to examine the
interface motion along the $z$-axis on a body-centered 
cubic (bcc) lattice.
Since it is much more convenient to implement the
latter case in higher dimensions we consider in this
work bcc lattices, the more as the lattice structure usually does not
affect the universal scaling behavior. 
\section{d=3}
\label{sec:d3}

\begin{figure}[t]
  \includegraphics[width=8.6cm,angle=0]{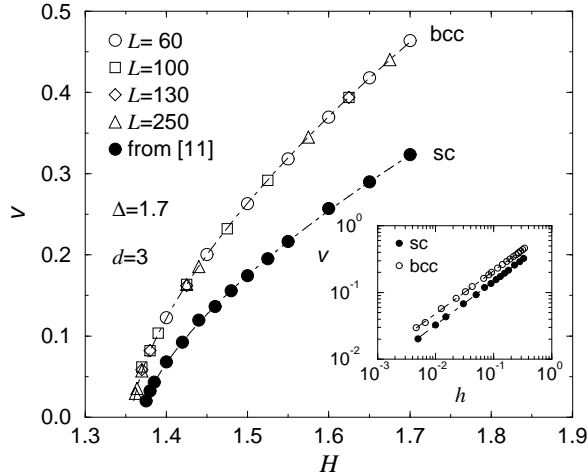}
  \caption{
    Dependence of the interface velocity $v$ on the driving field $H$
    for a bcc and simple cubic (sc) lattice, respectively.
    The inset shows $v$ as a
    function of the reduced driving field~$h$.
    The dash dotted lines are fits according to
    Eq.~(\protect\ref{eq:v_H_01}).  
   }
 \label{fig:bcc_sc_d3} 
\end{figure}

\begin{figure}[b]
  \includegraphics[width=8.6cm,angle=0]{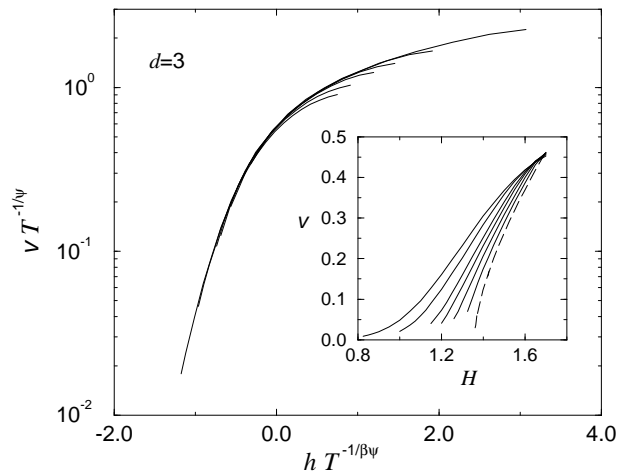}
  \caption{
    Scaling plot of the interface velocity for $d=3$. 
    The data are rescaled according to Eq.~(\ref{eq:tscal}).
    The inset shows the unscaled velocities for $T=0.025\,n$
    with $n \in \{1,2,3,4,6,8 \}$ (solid lines) in comparision to the
    $T=0$ data from Fig.~\ref{fig:bcc_sc_d3} (dashed line). 
    }
 \label{fig:d3_T_l0} 
\end{figure}
In the case of the three dimensional RFIM we consider
bcc lattices of linear size $L\le 250$.
A snapshot of a moving interface in the steady state is presented
in Fig.\,\ref{fig:snapshot}.
The obtained values of the interface velocities for $T=0$ 
are plotted in Fig.\,\ref{fig:bcc_sc_d3}.
As one can see $v$ tends
to zero in the vicinity of $H \approx 1.36$.
Assuming that the scaling behavior of the interface
motion is given by Eq.\,(\ref{eq:v_H_01})
one varies $H_{\text c}$ until one gets a straight line
in a log-log plot.
Convincing results are obtained for $H_{\text c}=1.357\pm0.001$
and the corresponding curve is shown in
the inset of Fig.\,\ref{fig:bcc_sc_d3}.
For lower and greater values of $H_{\text c}$ we observe
significant curvatures in the log-log plot (not shown).
In this way we estimate the error-bars in the determination
of the critical field.
A regression analysis yields the value
of the order parameter exponent $\beta=0.653\pm0.026$.
This value agrees with $\beta=0.66\pm0.04$ 
which was obtained from a similar investigation~\cite{ROTERS_1} 
where the interface moves along the diagonal direction of a
simple cubic lattice (see inset of Fig.\,\ref{fig:bcc_sc_d3}).
Furthermore, both results are in agreement
with $\beta=0.60\pm0.11$~\cite{AMARAL_1,AMARAL_2}, where in $d=3$ the
influence of helicoidal boundary conditions in one direction and
periodic ones in the other direction parallel to the interface was
investigated on a simple cubic lattice.

In order to determine the exponent $\psi$ we simulated
the RFIM around the critical field for
$T = 0.025\,n$ with $n \in \{ 1,2,3,4,6,8 \}$. 
The obtained curves are shown in the inset
of Fig.\,\ref{fig:d3_T_l0}.
According to Eq.\,(\ref{eq:v_scal_01})
the interface velocity scales as
\begin{equation}
  v(h,T)
  =
  T^{1/\psi} \,
  {\tilde v}(  h \,  T^{-1/\beta \psi}, 1 )  \, .
  \label{eq:tscal}
\end{equation}
Plotting $v(h,T)\, T^{-1/\psi}$ as a function of 
$h \, T^{-1/\beta \psi}$ one varies $\beta$, $\psi$,
and $H_{\text c}$ until one gets a data collapse of the 
different curves.
Convincing data collapses are observed for $\beta=0.63\pm0.06$
$\psi=2.33\pm0.2$ and $\HC=1.360\pm0.01$ and the corresponding
curves are shown in Fig.\,\ref{fig:d3_T_l0}.
The obtained values of the order parameter exponent and of the
critical field agree within the error-bars with the
values of the $T=0$ analysis.
Furthermore our results are in agreement with similar
investigations on a simple cubic lattice ($\beta=0.63\pm0.07$ and
$\psi=2.38\pm0.2$, see~\cite{ROTERS_1}).

\begin{figure}[b]
  \includegraphics[width=8.6cm,angle=0]{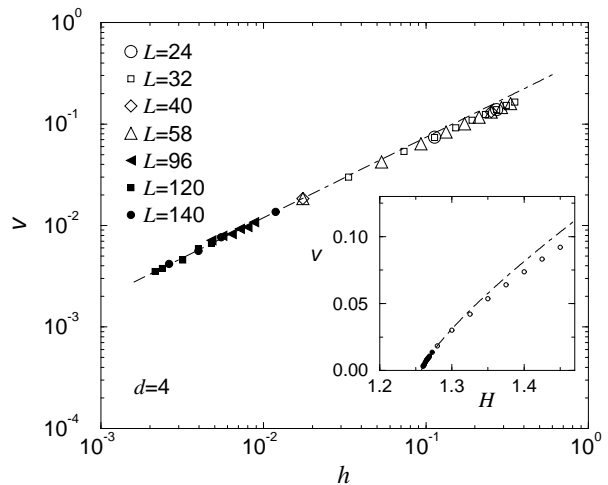}
  \caption{
    The interface velocity of the four-dimensional model at $T=0$. 
    For sufficiently small fields the data obey a power law
    according to Eq.\,(\ref{eq:v_H_01}) (dotted dashed line). 
    For the fit we use only those data marked by filled symbols
    and we find
    $H_\tindex{c}=1.258 \pm 0.05$
    and $\beta=0.8 \pm 0.06$. 
    }
 \label{fig:d4_T_eq0}
\end{figure}

\section{d=4}
\label{sec:d4}

\begin{figure}[t]
  \includegraphics[width=8.6cm,angle=0]{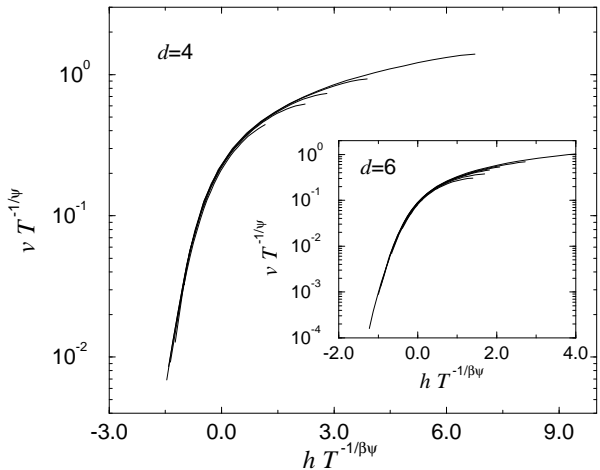}
  \caption{
    Scaling plot of the interface velocity for $d=4$ and $d=6$,
    respectively (see inset). 
    The data are rescaled according to Eq.~(\ref{eq:tscal}).
    }
 \label{fig:d4d6_T_l0} 
\end{figure}

In order to determine the order parameter exponent
of the four dimensional driven RFIM we measured the
interface velocity for bcc lattices of linear sizes
$L \le 140$.
The obtained data for $T=0$ are shown in a log-log plot in
Fig.\,\ref{fig:d4_T_eq0}.
After a transient regime which displays a finite curvature
we observe an asymptotic power-law behavior for sufficiently small
driving field~$H$.
A regression analysis yields 
$\beta=0.8 \pm 0.06$ and $H_\tindex{c}=1.258 \pm 0.002$.

To determine the exponent $\psi$ we simulated
the RFIM in the vicinity of the critical field for
$T = 0.025 \, n$ where again $n \in \{ 1,2,3,4,6,8 \}$ was choosen.  
Similar to the three dimensional case one varies the
exponents as well as the critical field until one observes a data
collapse. 
Good results are obtained for $\beta=0.73\pm 0.13$,
$\psi=1.72\pm 0.27$, and $H_{\text c}=1.256\pm0.015$.
The corresponding data collapse is shown in Fig.\,\ref{fig:d4d6_T_l0}.
Again, the obtained values of $\beta$ and $H_{\text c}$ confirm
the above presented analysis for $T=0$.
\section{d=5}
\label{sec:d5}

\begin{figure}[t]
  \includegraphics[width=8.6cm,angle=0]{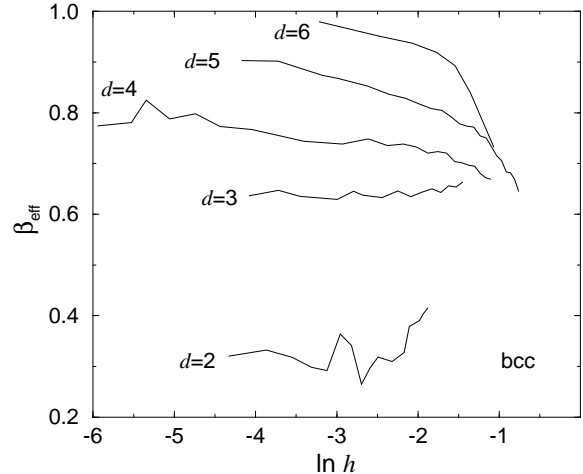}
  \caption{
    The effective exponents $\beta_{\rm eff}$ as a
    function of $\ln{h}$ for various dimensions.
    The figure shows that the five-dimensional
    exponent does not display a clear saturation as the
    exponent of the lower dimensions do.
    The data of the six-dimensional model indicate that the expected
    saturation value $\beta_{\text{eff}}=1$ is reached significantly
    faster as compared to $d=5$ (see text).%
    The values of the two-dimensional system
    are obtained from~\protect\cite{NOWAK_1}.
   }
 \label{fig:beta_eff} 
\end{figure}

In the case of the five-dimensional RFIM system sizes from $L=10$ up
to $L=30$ are simulated.
Analyzing the interface motion at $T=0$ we observe that the
velocity-field dependence can not be described by a pure power-law, 
i.e., Eq.\,(\ref{eq:v_H_01}) fails.
In Fig.~\ref{fig:beta_eff} we plotted the logarithmic derivation of
the velocity-field dependence 
\begin{equation}
  \beta_{\text{eff}} \; = \; 
  \frac{\partial\, \ln{v}}{\partial \, \ln{h}}
  \label{eq:beta_eff}
\end{equation}
which can be interpreted as an effective exponent.
If the asymptotic scaling behavior obeys Eq.\,(\ref{eq:v_H_01}) the
logarithmic derivative tends to the value of $\beta$
for $H \to H_{\text c}$.
But as can be seen from Fig.6
no clear saturation takes place for $d=5$ as it is observed for the
three and four dimensional model.
The lack of a clear saturation could be explained by a too large large
distance from the critical point, but another reason is possible too.

Significant deviations from a pure power law behavior
[Eq.\,(\ref{eq:v_H_01})] occur for instance 
at the upper critical
dimensional~$d_{\text c}$ where the scaling behavior is
governed by the mean-field exponents modified
by logarithmic corrections.
The scaling behavior around $d_{\text c}$ is
well understood within the renormalization 
group theory (see for instance~\cite{WEGNER_2,BREZIN_3,PFEUTY_1}).
For $d>d_{\text c}$ the stable fix point of the
corresponding renormalization equations is usually 
a trivial fix point with classical mean-field exponents.
This trivial fix point is unstable for $d<d_{\text c}$
and a different stable fix point exists
with nonclassical exponents.
These exponents can be estimated by an
$\epsilon$-expansion, for instance.
For $d=d_{\text c}$ both fix points are identical
and marginally stable.
In this case the asymptotic form of the thermodynamic functions is
given by the mean-field power-law behavior modified by logarithmic
corrections.
Applying this approach to the depinning transition, the corresponding
ansatz reads 
\begin{equation}
  v(h, T=0) \; \sim \; h^{\beta_\tindex{MF}}
  \; |\ln{h}|^{\text B},
  \label{eq:v_H_log}
\end{equation}
where~$\text B$ denotes an unknown correction exponent.
It is worth to mention that in contrast to the values of 
the critical exponents below the upper critical dimension 
the above scaling behavior
does not rely on approximation schemes like
$\epsilon$- or $1/n$-expansions~\cite{BREZIN_2}.
Within the renormalization group theory it is an exact result in the
limit $h\to 0$ (see e.g.~\cite{WEGNER_2,WEGNER_1} and references
therein for RG investigations and~\cite{GRIFFIN_1,BRINKMANN_1} for
measurements).

\begin{figure}[t]
  \includegraphics[width=8.6cm,angle=0]{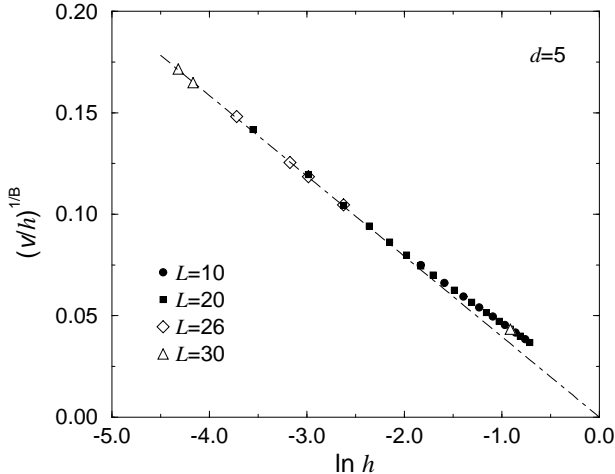}
  \caption{
    The rescaled interface velocity as a function of the
    driving field for $d=5$.
    In order to display the logarithmic corrections we plot
    $(v/h)^{1/{\text B}}$ vs.~$-\ln{h}$ [see Eq.\,(\ref{eq:v_H_log})].
    The solid line corresponds to the expected asymptotic scaling
    behavior for $h\to0$ [corresponding to $\ln h\to (-\infty)$]
    according to Eq.\,(\ref{eq:v_H_log_plot}).
   }
 \label{fig:d5_log_corr} 
\end{figure}

\begin{figure}[b]
  \includegraphics[width=8.6cm,angle=0]{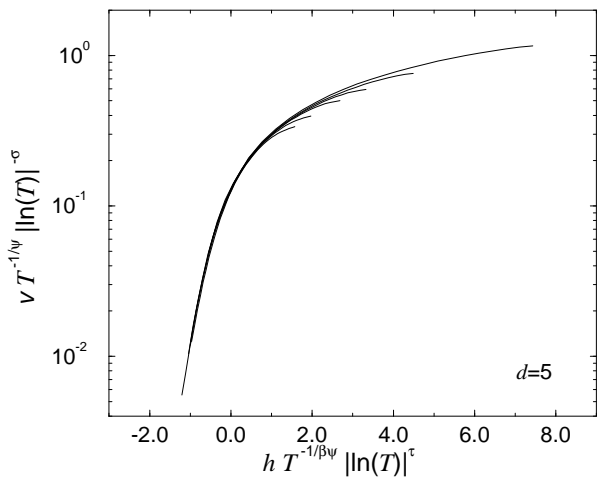}
  \caption{
    The scaling plot of the interface velocity~$v$ for the five
    dimensional model, i.e.\ at the upper critical dimension.
    The data are rescaled according to
    Eqs.\,(\ref{eq:ord_par_scal_dc},\ref{eq:ord_par_scal_arg_dc})
    using $\psi=1.49$, $\beta=1$, $\sigma=0.22$, and $\tau=0.19$.
    }
  \label{fig:d5_T_l0} 
\end{figure}

The value $\beta_\tindex{MF}=1$ is reported for depinning
transitions~\cite{KARDAR_2,NATTERMANN_1,CHAUVE_3}. 
Thus we analyze $v(h) /h$ as a function 
of $| \ln{h} |$ and note
again that Eq.\,(\ref{eq:v_H_log}) describes only 
the leading order of the scaling behavior, i.e., we 
expect that the asymptotic behavior of interface 
velocity obeys  
\begin{equation}
[ v(h)/h ]^{1/{\text B}} \; = \; {\rm const} \; |\ln{h} |.
\label{eq:v_H_log_plot}
\end{equation}
Therefore, we varied in our analysis the logarithmic
correction exponent~${\text B}$ and the critical field
$H_{\text c}$ until we get this expected asymptotic
behavior.
The best results are obtained for ${\text B}=0.40\pm 0.09$
and $H_{\text c} = 1.14235\pm 0.001$ and the corresponding
scaling plot is shown in Fig.\,\ref{fig:d5_log_corr}.
The observed asymptotic agreement with Eq.~(\ref{eq:v_H_log_plot})
corresponds to a logarithmically $(1/|\ln{h}|)$ convergence of
$\beta_\tindex{eff}$ to 
$\beta_\tindex{MF}=1$, which explains why no clear saturation of 
the effective exponent could be observed for $h\to 0$ in the five
dimensional model.

Similar to the $T=0$ scaling behavior one has to modify for 
$T>0$ the scaling ansatz since no data collapse could 
be obtained by plotting the data according 
to Eq.\,(\ref{eq:v_scal_01}).
Motivated by recently performed investigations of the
scaling behavior of an absorbing phase transition
around the upper critical dimension~\cite{LUEB_22}
we assume that the scaling behavior of the order parameter
obeys in leading order 
\begin{equation}
v(h,T) \; = \; T^{1/\psi_{\tindex{MF}}} \, |\ln{T}|^{\sigma}
\; {\tilde v}(x,1)
\label{eq:ord_par_scal_dc}
\end{equation}
where the scaling argument~$x$ is given in leading order by
\begin{equation}
x \; = \; h\, T^{-1/\beta_{\tindex{MF}}\psi_{\tindex{MF}}} \,
|\ln{T}|^{\tau}
\label{eq:ord_par_scal_arg_dc}
\end{equation}
with $\beta_{\tindex{MF}}=1$.
In our analysis we use the value $\psi_{\rm MF}=1.49$ obtained from
the analysis of the six dimensional RFIM (see next section). 
Therefore, we have to vary the exponents 
$\sigma$ and $\tau$ in order to observe a data collapse according to
Eqs.\,(\ref{eq:ord_par_scal_dc},\ref{eq:ord_par_scal_arg_dc}).
Convincing results are obtained for $\sigma=0.22\pm0.16$,
$\tau=0.19\pm0.12$
The corresponding data collapse is shown in Fig.\,\ref{fig:d5_T_l0}.

\section{d=6}
\label{sec:d6}

\begin{figure}[t]
  \includegraphics[width=8.6cm,angle=0]{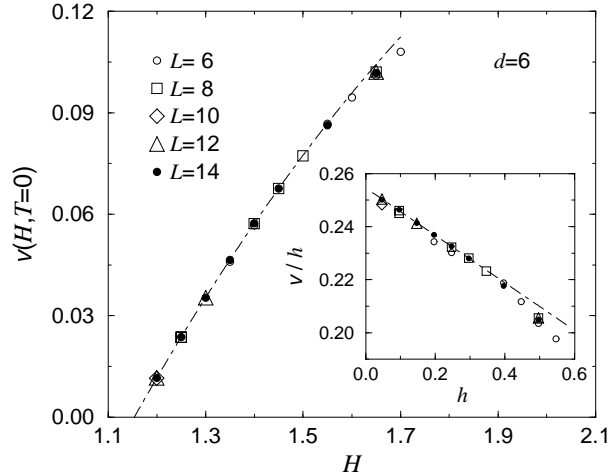}
  \caption{
    The interface velocity as a function of the
    driving field for the six-dimensional model.
    The dashed line corresponds to a fit according to 
    Eq.\,(\protect\ref{eq:v_H_d6_02}).
    The inset displays $v(h)/h$ as a function of the 
    reduced driving field $h$.
    The resulting linear behavior confirms that the deviations from
    the mean-field behavior can be described by quadratic
    corrections. 
   }
 \label{fig:v_H_d6} 
\end{figure}

Above the upper critical dimension the 
scaling behavior is characterized by the 
mean-field exponents, i.e., in
leading order the interface velocity is given by 
\begin{equation}
v(h, T=0) \; \sim \; \, h. 
\label{eq:v_H_d6_01}
\end{equation}
In Fig.\,\ref{fig:v_H_d6} we plot the velocity 
as a function of the driving field~$H$ obtained
from simulations of system sizes $L\le 14$.
As one can see the velocity does not display
the expected linear behavior.
It seems that a linear behavior is only given
for small velocities, i.e., our data do not
display the pure asymptotic behavior [Eq.\,(\ref{eq:v_H_d6_01})].
This is confirmed by the behavior of the effective
exponent [Eq.(\ref{eq:beta_eff})] which increases fast for $h\to 0$
but the actual saturation to $\beta=1$ does not
take place for the considered values of $h$
(see Fig.\,\ref{fig:beta_eff}).
To observe the asymptotic behavior one has to
perform simulations closer to the critical point $H_{\text c}$
which requires to simulate larger system sizes.
Unfortunately, the limited CPU power makes this impossible.

An alternative is to take the curvature of the function $v(h)$
into consideration and to assume that the leading
corrections to the asymptotic behavior are of the form
\begin{equation}
v(h,T=0) \; = \; v_1 \, h \; + \;
v_2 \, h^2 \; + \; {\cal O}(h^3),
\label{eq:v_H_d6_02}
\end{equation}
which recovers Eq.~(\ref{eq:v_H_d6_01}) for $h\to 0$.
Fitting our data to this ansatz we
get $H_{\text c}=1.1537\pm 0.003$, $v_1=0.2546$,
and $v_2=-0.0895$.
The corresponding curve fits the simulation data quite
well as one can see from Fig.\,\ref{fig:v_H_d6}.
In the inset of Fig.\,\ref{fig:v_H_d6} we plotted 
$v(h)/h$ as a function of the reduced driving 
field~$h$.
According to the above ansatz~[Eq.\,(\ref{eq:v_H_d6_02})] one
gets a linear behavior, i.e., the deviations from
the pure mean-field behavior [Eq.\,(\ref{eq:v_H_d6_01})]
can really be described as quadratic corrections.
Thus we get that our numerical data are consistent with
the assumption that the six dimensional RFIM depinning
transition is characterized by the mean-field 
exponent $\beta_{\tindex{MF}}=1$.

Again we consider how thermal fluctuations affect the scaling
behavior and analyse interface velocities obtained at different
temperatures $T=0.025 \, n$ with 
$n \in \{ 1,2,3,4,6,8 \}$. 
Similar to the situation below the upper critical dimension we 
assume that the scaling behavior of the interface velocity
is given by Eq.\,(\ref{eq:v_scal_01}) where the exponents 
are given by mean-field values. 
A convincing data collapse is obtained for 
$\psi_{\tindex{MF}}=1.49\pm0.15$ and $\HC=1.153\pm0.02$
and is plotted in the inset of Fig.~\ref{fig:d4d6_T_l0}.

\section{Discussion}
\label{sec:dis}

\begin{figure}[t]
  \includegraphics[width=8.6cm,angle=0]{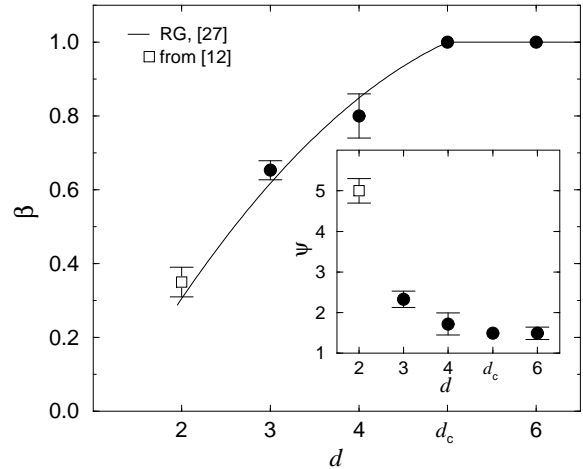}
  \caption{
    The critical exponents $\beta$ and $\psi$
    as a function of the dimension.
    The solid line corresponds to the values of an
    $\epsilon$-expansion [Eq.~(\ref{eq:beta_chauve})].
    }
 \label{fig:exp_dim}
\end{figure}

A well established realization
of interface pinning in a disordered media is the
so-called quenched Edwards-Wilkinson (QEW) equation of
motion which was intensively investigated in the last
decade~\cite{BRUINSMA,NARAYAN_1,NATTERMANN_1,CHAUVE_1,CHAUVE_2,CHAUVE_3}.
It is argued that the QEW equation as well as the driven RFIM
are characterized by the same critical exponents,
i.e., both models belong to the same universality 
class~\cite{BRUINSMA,AMARAL_1}.
Renormalization group analyses of the quenched QEW
equation~\cite{NATTERMANN_1,CHAUVE_3} predict, in accordance
with~\cite{BRUINSMA},  
$d_{\text c}=5$ and allow to estimate the critical exponents 
using an $\epsilon$-expansion.
A recently performed two-loop renormalization approach 
yields~\cite{CHAUVE_3}
\begin{equation}
 \beta_{\text{QEW}}\; = \; 1-\frac{1}{9}\,\epsilon-0.040123\,\epsilon^2
 + {\cal O}(\epsilon^3)
  \label{eq:beta_chauve}
\end{equation}
where $\epsilon$ denotes the distance from the upper critical
dimension, i.e.\ $\epsilon=5-d$ (unfortunately, no error-bars can 
be estimated from an $\epsilon$-expansion).
The corresponding values of the exponents as a function
of the dimension are plotted in Fig.~\ref{fig:exp_dim}.
The numerically determined exponents~$\beta$ of the driven 
RFIM (listed in Table~\ref{table:exponents}) are in a 
fair agreement with the values of the $\epsilon$-expansion.

For the QEW equation the temperature exponent $\psi$ is 
not known.
Therefore, a direct comparison with the obtained values of the
driven RFIM is not possible.

\begin{table}[b]
  \caption{%
    The exponents $\beta$
    (obtained from simulations at $T=0$ and $T>0$, respectively)
    and $\psi$
    of the depinning transition of the RFIM for different dimensions. 
    The values of the two dimensional model are obtained 
    from~\protect\cite{NOWAK_1}.
    The critical behavior at the upper critical dimension~$d_\tindex{c}$
    is additionally affected by logarithmic corrections. 
    }
  \label{table:exponents}
  \begin{tabular}{llll}
    $d$        & $\beta_{T=0}$   &  $\beta_{T>0}$ & $\psi$\\  
    \colrule \\
    $2$             & $0.35\pm 0.04$          &  $0.33\pm0.02\quad$ & $5.00  \pm 0.3$\\ 
    $3$             & $0.653\pm 0.026$\quad$$ &  $0.63\pm0.06$      & $2.33 \pm 0.2$\\ 
    $4$             & $0.80\pm 0.06$          &  $0.73\pm0.13$      & $1.72 \pm 0.27$\\ 
    $d_\tindex{c}=5\quad$& $1$                &  $1$                & $1.49$          \\  
    $6$\quad        & $1$                     &  $1$                & $1.49\pm 0.15$  \\       
  \end{tabular}
\end{table}
\section{Conclusions}
\label{sec:conc}

We studied numerically a field driven interface in the RFIM 
and determined the order parameter exponent $\beta$ as
well as the temperature exponent $\psi$.
Below the upper critical dimension $d_{\text c}=5$ 
the critical exponents depend on the dimension and the 
values of the exponents correspond to those of a two-loop 
renormalization group approach of the 
Edwards-Wilkinson equation~\cite{CHAUVE_3}.
This suggest that the depinning transition of the RFIM model belongs
to the universality class of the quenched Edward-Wilkinson equation.
At the upper critical dimension $d_{\text c}=5$ the scaling 
behavior is affected by logarithmical corrections.
Above the upper critical dimension we observe that 
the scaling behavior is characterized in leading order 
by the corresponding mean-field exponents.

\acknowledgments
This work was supported by the Deutsche Forschungsgemeinschaft via
GRK~277
{\it Struktur und Dynamik heterogener Systeme} 
(University of Duisburg)
and SFB~491
{\it Mag\-ne\-tische Hetero\-schich\-ten: Struktur und el\-ek\-tronischer
  Transport}
(Duisburg/Bochum).

S.\,L\"ubeck wishes to thank the Minerva 
Foundation (Max Planck Gesellschaft) for financial support.


\end{document}